\documentclass[aps,prd,superscriptaddress,preprint,tightenlines,nofootinbib,floatfix]{revtex4}

\usepackage{graphicx} % Include figure files
\usepackage{dcolumn}  % Align table columns on decimal point
\usepackage{bm}       % bold math

\begin{document}

%\preprint line(s) will be ignored for PRL/PRD
\preprint{CLNS-2058} % For paper draft CBX YY-NN -> Draft YY-NNA

% Add \boldmath if you DO have mathematical symbols in the title
\title{\boldmath Search for the Decay $J/\psi\rightarrow\gamma + invisible$}

%-------- INSERT HERE ------------
% Your author list goes here  REMOVE EVERYTHING to END INSERT and
% replace with your authorlist (ask cleoac).

\author{J.~Insler}
\author{H.~Muramatsu}
\author{C.~S.~Park}
\author{E.~H.~Thorndike}
\author{F.~Yang}
\affiliation{University of Rochester, Rochester, New York 14627, USA}
\author{S.~Ricciardi}
\affiliation{STFC Rutherford Appleton Laboratory, Chilton, Didcot, Oxfordshire, OX11 0QX, UK}
\author{C.~Thomas}
\affiliation{STFC Rutherford Appleton Laboratory, Chilton, Didcot, Oxfordshire, OX11 0QX, UK}
\affiliation{University of Oxford, Oxford OX1 3RH, UK}
\author{M.~Artuso}
\author{S.~Blusk}
\author{S.~Khalil}
\author{R.~Mountain}
\author{T.~Skwarnicki}
\author{S.~Stone}
\author{J.~C.~Wang}
\author{L.~M.~Zhang}
\affiliation{Syracuse University, Syracuse, New York 13244, USA}
\author{G.~Bonvicini}
\author{D.~Cinabro}
\author{A.~Lincoln}
\author{M.~J.~Smith}
\author{P.~Zhou}
\author{J.~Zhu}
\affiliation{Wayne State University, Detroit, Michigan 48202, USA}
\author{P.~Naik}
\author{J.~Rademacker}
\affiliation{University of Bristol, Bristol BS8 1TL, UK}
\author{D.~M.~Asner}
\author{K.~W.~Edwards}
\author{K.~Randrianarivony}
\author{J.~Reed}
\author{A.~N.~Robichaud}
\author{G.~Tatishvili}
\author{E.~J.~White}
\affiliation{Carleton University, Ottawa, Ontario, Canada K1S 5B6}
\author{R.~A.~Briere}
\author{H.~Vogel}
\affiliation{Carnegie Mellon University, Pittsburgh, Pennsylvania 15213, USA}
\author{P.~U.~E.~Onyisi}
\author{J.~L.~Rosner}
\affiliation{University of Chicago, Chicago, Illinois 60637, USA}
\author{J.~P.~Alexander}
\author{D.~G.~Cassel}
\author{R.~Ehrlich}
\author{L.~Fields}
\author{L.~Gibbons}
\author{S.~W.~Gray}
\author{D.~L.~Hartill}
\author{B.~K.~Heltsley}
\author{J.~M.~Hunt}
\author{D.~L.~Kreinick}
\author{V.~E.~Kuznetsov}
\author{J.~Ledoux}
\author{H.~Mahlke-Kr\"uger}
\author{J.~R.~Patterson}
\author{D.~Peterson}
\author{D.~Riley}
\author{A.~Ryd}
\author{A.~J.~Sadoff}
\author{X.~Shi}
\author{S.~Stroiney}
\author{W.~M.~Sun}
\affiliation{Cornell University, Ithaca, New York 14853, USA}
\author{J.~Yelton}
\affiliation{University of Florida, Gainesville, Florida 32611, USA}
\author{P.~Rubin}
\affiliation{George Mason University, Fairfax, Virginia 22030, USA}
\author{N.~Lowrey}
\author{S.~Mehrabyan}
\author{M.~Selen}
\author{J.~Wiss}
\affiliation{University of Illinois, Urbana-Champaign, Illinois 61801, USA}
\author{M.~Kornicer}
\author{R.~E.~Mitchell}
\author{M.~R.~Shepherd}
\author{C.~M.~Tarbert}
\affiliation{Indiana University, Bloomington, Indiana 47405, USA }
\author{D.~Besson}
\affiliation{University of Kansas, Lawrence, Kansas 66045, USA}
\author{T.~K.~Pedlar}
\author{J.~Xavier}
\affiliation{Luther College, Decorah, Iowa 52101, USA}
\author{D.~Cronin-Hennessy}
\author{K.~Y.~Gao}
\author{J.~Hietala}
\author{R.~Poling}
\author{P.~Zweber}
\affiliation{University of Minnesota, Minneapolis, Minnesota 55455, USA}
\author{S.~Dobbs}
\author{Z.~Metreveli}
\author{K.~K.~Seth}
\author{X.~Ting}
\author{A.~Tomaradze}
\affiliation{Northwestern University, Evanston, Illinois 60208, USA}
\author{S.~Brisbane}
\author{J.~Libby}
\author{L.~Martin}
\author{A.~Powell}
\author{P.~Spradlin}
\author{G.~Wilkinson}
\affiliation{University of Oxford, Oxford OX1 3RH, UK}
\author{H.~Mendez}
\affiliation{University of Puerto Rico, Mayaguez, Puerto Rico 00681}
\author{J.~Y.~Ge}
\author{D.~H.~Miller}
\author{I.~P.~J.~Shipsey}
\author{B.~Xin}
\affiliation{Purdue University, West Lafayette, Indiana 47907, USA}
\author{G.~S.~Adams}
\author{D.~Hu}
\author{B.~Moziak}
\author{J.~Napolitano}
\affiliation{Rensselaer Polytechnic Institute, Troy, New York 12180, USA}
\author{K.~M.~Ecklund}
\affiliation{Rice University, Houston, Texas 77005, USA}
\collaboration{CLEO Collaboration}
\noaffiliation

%-------- END INSERT ------------

%please hard code the date when you have a final draft and submit to CLEOAC
\date{\today}

\begin{abstract} 
A search for $J/\psi$ radiative decay to weakly interacting neutral final states was performed using the CLEO-c detector at CESR.   $J/\psi$ events were selected by observing the hadronic decay $\psi(2S)\rightarrow\pi^{+}\pi^{-}J/\psi $.
A total of $3.7\times 10^{6}$ $J/\psi$ events was used to study the decay $J/\psi\rightarrow\gamma+X$, where $X$ is a narrow state that is invisible to the detector.  No significant signal was observed and upper limits on the branching fraction were set for masses $m_X$ up to 960 MeV/$c^2$.  
The upper limit corresponding to $m_X=0$  is $4.3\times 10^{-6}$ at the $90\%$ confidence level.
\end{abstract}

\pacs{13.20.Gd, 13.66.Hk, 14.80.Cp, 14.80.Ly}
\maketitle

Understanding the nature of dark matter is one of the major goals of particle physics.  In one scenario two dark-matter particles annihilate in the early universe primarily through the s-channel production of a light neutral boson that lies outside the standard model.  Coupling strengths can be chosen that  allow both the dark-matter particle and the boson to decouple from standard-model particles and thus avoid conflict with experimental limits on their laboratory production rates,  while at the same time giving the correct remnant dark-matter density~\cite{Larios:2002ha, Anderson:2003bj, Fayet:2006sp, Fayet:2007ua, Borodatchenkova:2005ct, Hooper:2008im, Gunion:2005rw, Boehm:2003bt}.  In these models the lightest neutralino is a logical candidate for the dark-matter particle but a light scalar is also a possibility.  Non-minimal supersymmetric models that include a scalar singlet Higgs field produce a rich spectrum of Higgs and neutralino particles that can extend to very low masses~\cite{Barger:2006dh, Barger:2005hb} so they can be easily applied to the dark-matter problem.   
For example, the required light boson can be a gauge 
boson~\cite{Fayet:2007ua, Borodatchenkova:2005ct, Hooper:2008im, Boehm:2003bt} or a CP-odd Higgs boson~\cite{Gunion:2005rw, Larios:2002ha}.

Experimental limits on the branching fractions for quarkonium radiative decay to invisible final states~\cite{Balest:1994ch, Besson:1985xw} provide laboratory constraints for some of the models discussed above~\cite{Fayet:2007ua, Gunion:2005rw}.  Preliminary data on $\Upsilon(3S)\rightarrow\gamma+invisible$ have also been reported~\cite{Aubert:2008st}.  These can be incorporated into theoretical models by calculating the branching fraction for a light boson to be produced in radiative quarkonium decay and then assuming its dominant decay is to two neutralinos.  The neutralinos interact very weakly with normal matter so they would not be detected in an experiment.    In this case the charmonium radiative decay amplitude takes the form,

\begin{equation}
\label{eq:axion}
\frac{\Gamma (J/\psi \rightarrow \gamma X)}{\Gamma (J/\psi \rightarrow \mu^{+}\mu^{-} )}\approx \frac{G_{F}m_{c}^{2}}{2\sqrt{2}\alpha \pi} \frac{\cos ^{2}\theta}{\tan ^{2}\beta}
\end{equation}

\noindent 
where $X$ is a boson with pseudoscalar quark coupling, $\theta$ is the Higgs mixing angle, $m_{c}$ is half the $J/\psi$ mass, and $\tan\beta$ is the usual ratio of vacuum expectation values.  A factor of $1/2$ is included in this expression as a rough estimate of radiative and relativistic corrections~\cite{Fayet:2007ua}.  The corresponding expression for bottomonium decay takes the same form but with $\tan ^{2}\beta$ appearing in the numerator instead of the denominator (Eq.~(111) in Ref.~\cite{Fayet:2007ua}).

In this letter we report upper limits on the branching fraction for $J/\psi\rightarrow\gamma+invisible$.  Here an invisible particle is one  that escapes detection because its interaction with the detector is very weak, or because it decays to other neutral weakly-interacting particles within the detector volume.  The present data provide new constraints for some models in which a light boson decays to two dark-matter particles.

Data were acquired at the $\psi(2S)$ mass using the Cornell Electron Storage Ring (CESR) with the CLEO-c detector~\cite{Viehhauser:2001ue, Peterson:2002sk, Artuso:2005dc}.  Photons were detected in a CsI(Tl) electromagnetic calorimeter, which has photon energy resolution equal to 2.2$\% $ at 1 GeV and 5$\% $ at 100~{ MeV}.  For measurements in the charm region the CLEO~III silicon vertex detector~\cite{Coan:1999zk} was replaced with a cylindrical drift-chamber and the solenoid magnetic field was set to 1.0~{ T}.  

\begin{figure}
\includegraphics[width=8.6cm]{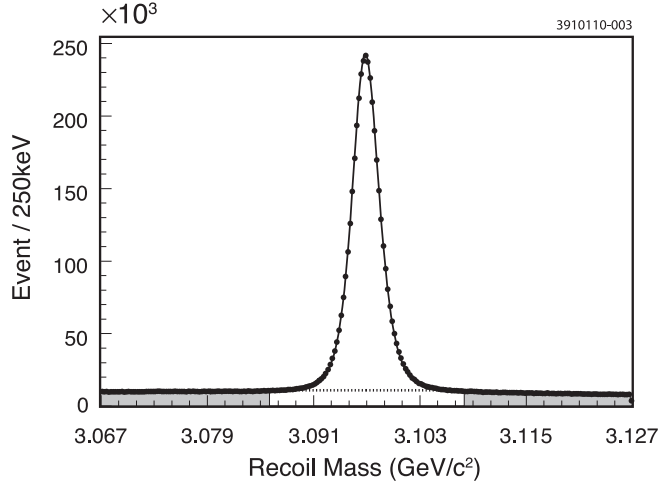}%
\caption{\label{recoil}Recoil mass calculated from the di-pion four-momentum, after pion selection.  A fit to the unshaded region of the spectrum was used to extract the tagged $J/\psi$ yield.  The solid line shows the results of the fit and the dotted line shows the background contribution.  Data in the shaded regions were used to study background.}
\end{figure}

$J/\psi $ events were tagged by measuring the charged pions from  $\psi(2S)\rightarrow\pi^{+}\pi^{-}J/\psi $.  All events that had additional charged tracks beyond the tagging pions were rejected.  
Event selection criteria were chosen to optimize the identification of charged pions and minimize the background from hadronic showers produced in the calorimeter.  Transition pions were required to have differential energy loss signatures consistent with a pion and their trajectories were required to form a single decay vertex close to the $e^{+}e^{-}$ interaction point.   In addition, individual pion momentum vectors were required to lie in the central region of the detector with $\left|\cos{\theta^{\prime}}\right|<0.83$, where $\theta^{\prime}$ is measured relative to the beam axis, and the summed momentum vector for each pion pair was required to have $\left|\cos{\theta^{\prime}}\right|<0.95$.  The latter constraint was used to suppress background from  $\gamma \gamma$ fusion and direct $J/\psi$ production from $e^{+}e^{-}\rightarrow\gamma J/\psi$.   Figure~\ref{recoil} shows the recoiling invariant mass calculated from the four-momentum of the tagging pions and the center-of-mass energy of the initial $e^{+}e^{-}$ system.  This recoil mass was  required to be within $\pm$5~MeV/$c^{2}$ of the $J/\psi $ mass for the selected invisible-decay candidates.
Further background reduction was achieved by requiring the invariant mass of the $\pi^{+}\pi^{-} $ pair to be between 460 and 590 MeV/$c^{2}$, and requiring charged tracks to have momentum component transverse to the beam in excess of 100 MeV/c. Those criteria were determined by maximizing $S^{2}/B$, where $S$ is the inclusive signal yield in the $J/\psi $ mass peak and $B$ is the scaled background yield in dipion recoil mass side-band regions (Fig.~\ref{recoil}).  The tagged $J/\psi $ yield, 
$3.7\times 10^{6}$, was determined by fitting the data in Fig.~\ref{recoil} to a sum of three Gaussians and a third-order polynomial background function.

The most energetic calorimeter shower in each event that was not associated with a 
transition-pion track was designated as the signal photon corresponding to $J/\psi $ radiative decay. 
This signal photon must lie in the barrel region of the 
detector, with $\left|\cos{\theta^{\prime}}\right|<0.79$, and when transformed 
into the $J/\psi $ rest frame it must have energy exceeding 1.25 GeV.  The latter selection avoided the rapidly rising background below 1.25 GeV.  The photon shower was also required to have a lateral shape consistent with that 
expected for a photon so as to suppress anti-neutron showers 
from the decay $J/\psi\rightarrow n\bar{n}$.

Because the signal-event topology includes only transition pions and a signal photon visible 
in the detector, restricting additional calorimeter 
activity should in principle only aid in rejecting 
background from other $J/\psi $ decays and not affect the  
invisible signal. However the transition pions 
themselves interact in the calorimeter and produce 
shower fragments. Hence care was 
taken so as not to reject events in which such 
remnants of hadronic interactions are produced, 
while still eliminating background events with additional photons. 
Showers were considered as photon candidates 
only if they were not in close geometrical proximity 
to the calorimeter entry point of a transition pion,  
and if they survived both a lateral-profile 
selection and the application of a neural network. The lateral-profile restriction eliminated broad 
showers from consideration, and the neural network 
examined the energy distribution among the 
crystals in a shower. (See Fig. 12 in Ref.~\cite{Gibbons:1998}.)
Events were removed from consideration if there were any additional 
photon candidates with energy greater than 50 MeV. 
This restriction was found to reject just 
$1\%$ of a test sample selected by requiring $J/\psi\rightarrow\mu^{+}\mu^{-}$ decay instead of the signal-photon requirement.

In addition to the photon selection criteria discussed above, a final selection was made on the total energy deposited in the calorimeter minus the signal-photon energy.  Events above 600 MeV were rejected, which reduced the background from 
events in which one or more energetic photons coincidentally 
overlapped the shower caused by one of the transition pions.  This removed much of the high-energy background in the signal-photon spectrum arising from radiative $J/\psi$ decay to $\eta$ and $\eta^{\prime}$.  The limit was set by comparing the results of Monte Carlo simulations for these two channels and for radiative decay to a massless invisible particle.  The signal-photon selection efficiency, $\epsilon$, fell from about $65\%$ to $55\%$ after this data selection was made.  

\begin{figure}
\includegraphics[width=8.6cm]{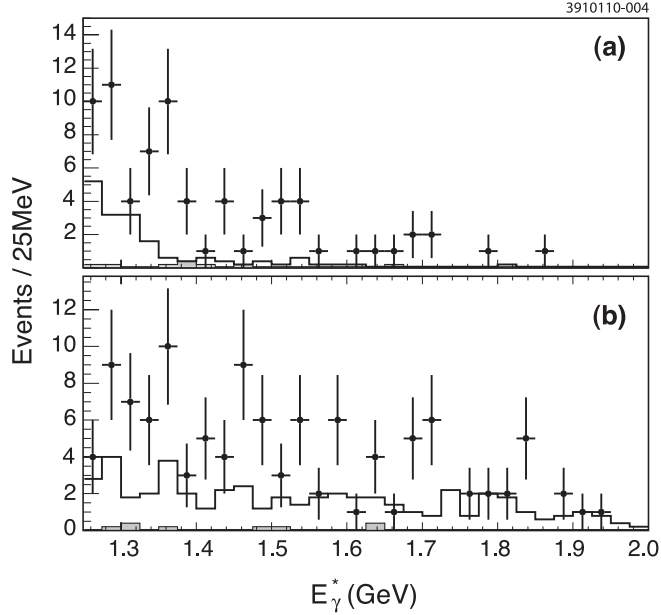}%
\caption{\label{spectrum}(a) Invisible-decay events as a function of photon energy in the $J/\psi$ rest frame, and (b) the anti-neutron-enhanced spectrum obtained by selecting calorimeter showers with large lateral extent.  The solid points are data and the unshaded histograms show the results of Monte Carlo simulations.  The shaded histograms show the small contributions from events that have di-pion recoil mass adjacent to the $J/\psi$ mass window. }
\end{figure}

Figure~\ref{spectrum}(a) shows a histogram of the selected data for $J/\psi\rightarrow\gamma + invisible$ as a function of $E_{\gamma}^{*}$, the photon energy in the $J/\psi$ rest frame.   A total of 73 data events were measured with $E_{\gamma}^{*}\geq 1.25$ GeV.   A sharp peak in this spectrum would be evidence for $J/\psi $ radiative decay to a narrow final state.  No obvious peak corresponding to $J/\psi\rightarrow\gamma + invisible$ is apparent in the data.    A Monte Carlo simulation of the background spectrum, normalized to integrated beam luminosity, is also shown.  

To study the background from channels such as $\psi(2S)$ decay, data were extracted from recoil-mass regions adjacent to the $J/\psi$ (Fig.~\ref{recoil}).   In Fig.~\ref{spectrum} this  sideband yield has been 
scaled down by the ratio of the mass intervals used to select the on-resonance and sideband data. The on-resonance data 
are seen to originate predominantly from $J/\psi$ decays,
as the sideband yield is negligibly small. 

Although there
are no obvious peaks in the on-resonance data, there does appear to be a
relatively smooth background, the size of
which is underestimated by the Monte Carlo simulation. This background has the
somewhat surprising property that it continues well above the
photon kinematic limit of $E_{\gamma}^{*}=1.55$ GeV, meaning it cannot be solely
from radiative $J/\psi$ decays.  Monte Carlo studies indicate that the dominant background
contribution in Fig.~\ref{spectrum}(a), in fact the only one for
$E_{\gamma}^{*}\geq 1.4$ GeV, is from $J/\psi\rightarrow n\bar{n}$ decay.
These events have the neutron leaving no detectable
signature in the tracking or shower detectors,
and the anti-neutron producing a shower in the calorimeter that is misidentified as a high-energy photon. 
Since anti-neutrons annihilate in the calorimeter they can produce shower energies that exceed
the kinematic limit for photons by converting part of a nucleon rest mass into visible energy.
This background source was studied 
by selecting calorimeter showers with large lateral spread, 
effectively eliminating real photon showers and
preferentially selecting background events that did not have a photon in the final state. The $E_{\gamma}^{*}$
distribution of these events appears in
Fig.~\ref{spectrum}(b) for data and Monte Carlo, where two
salient features can be observed: it shows no cutoff at the
kinematic limit for photons, just as in Fig.~\ref{spectrum}(a), 
and it is a smooth function of $E_{\gamma}^{*}$. The GEANT3 Monte Carlo simulation~\cite{GEANT:1993} correctly models the spectral shape but not the absolute number of anti-neutrons that
satisfy the photon-like lateral-shower criterion.  However since this is a very small
fraction of all anti-neutrons, it is not unexpected.

In order to extract 
upper limits on the branching fraction for $J/\psi\rightarrow\gamma + invisible$ a series of 
fits was made to the selected on-resonance data.
Two-body radiative $J/\psi$ decay to a narrow final state features a monochromatic photon in the $J/\psi$ center-of-mass frame.  This produces a peak in the measured $E_{\gamma}^{*}$ spectrum with a line shape determined by the calorimeter resolution.  We extracted the branching fraction for $J/\psi\to\gamma X$ for fixed mass, $m_X$, with a binned maximum likelihood fit to the $E_{\gamma}^{*}$ spectrum.   The fit assumed an exponentially falling background and a peak lineshape determined from a Monte Carlo simulation of the detector response to signal events.  The peak lineshape was parameterized with a Crystal Ball function~\cite{Gaiser:1982, Skwarnicki:1986}, which features a low-side tail joined smoothly to a Gaussian core.  The $E_{\gamma}^{*}$ peak resolution was approximately 30 MeV.  

\begin{figure}
\includegraphics[width=8.6cm]{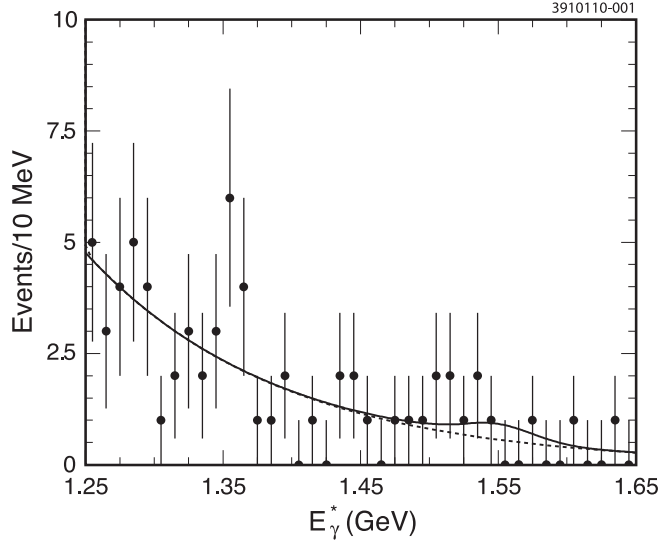}%
\caption{\label{fit}Fit to determine the branching fraction at $E_{\gamma}^{*}=1.548$~GeV, corresponding to $m_X=0$.  The data are the same as in Fig.~\ref{spectrum}(a) but with finer binning.  The solid line is the total fitted spectrum, and the dashed line shows the background contribution to the fit.}
\end{figure}

Fits were performed on the data in the $E_{\gamma}^{*}$ range from 1.25 to 1.65~GeV, with fixed peak energies chosen from 1.400~GeV to 1.548~GeV in 5 MeV steps.  A typical fit is shown in Fig.~\ref{fit}.  This region encompasses the kinematic threshold at 1.55~GeV and probes a broad range of light invisible-particle masses recoiling against the photon, up to 960~MeV/$c^2$. The data in this range are well described by an exponential background curve alone, showing no evidence for a signal.  We also note the possible presence of a narrow structure in the spectrum at about 1.36~GeV.  This peak is much narrower than the expected photon-energy resolution, and is therefore attributed to a statistical fluctuation in the data.  

The branching fraction for invisible decay was calculated as 
$B(J/\psi \rightarrow \gamma +invisible)=N_{inv}/(\epsilon N_{tag})$.
Here $N_{tag}$ is the number of tagged inclusive $J/\psi$ events, $N_{inv}$ is the number of fitted events after photon selection and particle vetoes, and $\epsilon$ is the efficiency for selecting invisible radiative decays from the tagged $J/\psi$ sample, as determined from Monte Carlo simulations. 

As there were no strong peaks in any of the fits, branching fraction upper limits were extracted at the $90\%$ confidence level at each $m_X$.  This was done by integrating the likelihood for positive branching fractions. The limits are dominated by statistical uncertainties but vary somewhat depending on the shape of the background function (exponential or polynomial), range of the fit, and reasonable variations in event selection criteria. These systematic variations were added in quadrature to estimate the systematic uncertainty in the fitted yields.  Systematic uncertainties were included in the branching fraction upper limits by scaling them upward by the same amount that Gaussian upper limits were  changed when systematic and statistical uncertainties were added in quadrature.

\begin{figure}
\includegraphics[width=8.6cm]{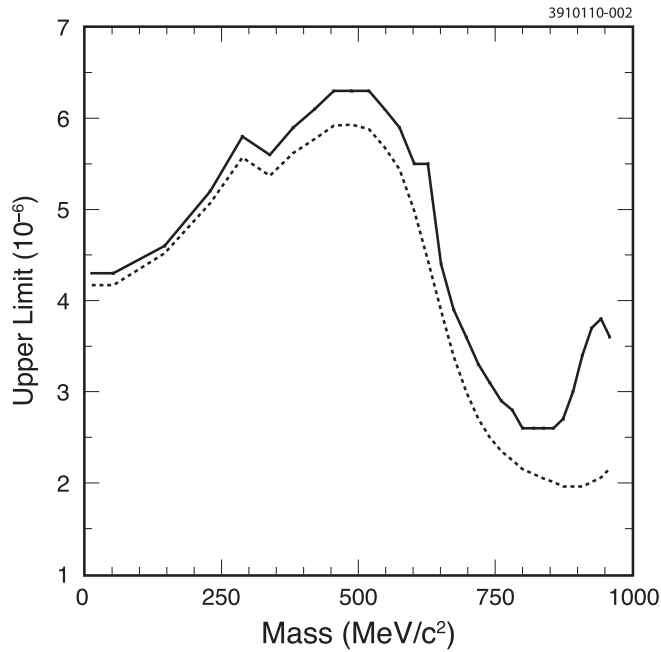}%
\caption{\label{limit}The $90\%$ confidence-level upper limits for $J/\psi\rightarrow\gamma X$, where $X$ is invisible to the detector.  The dashed line shows the results for statistical uncertainties alone, and the solid line includes systematic and statistical uncertainties.}
\end{figure}

The resulting $90\%$ confidence-level upper limits on the branching fraction are plotted in Fig.~\ref{limit} as the solid line.  The dashed line shows the upper limit for statistical uncertainties alone.  Statistical uncertainties are dominant for the full mass range in the plot, and systematic uncertainties are negligible for the low-mass points.
The upper limits from this experiment are well below the previous experimental limits for charmonium radiative decay to invisible final states~\cite{Edwards:1982zn}. 

The present data yield a branching fraction for $J/\psi\rightarrow~\gamma +invisible$ of $(1.5 \pm  2.4)\times 10^{-6}$ at $m_X=0$, and an upper limit of $4.3\times 10^{-6}$ at the $90\%$ confidence level.  The error quoted here ($2.4\times 10^{-6}$) is from the statistical error in the fitted peak only.  
Substituting this upper limit in Eq.~(1) gives 
$\cos^{2} \theta/\tan^{2} \beta \leq 0.084$.  When combined with experimental limits for $\Upsilon$ radiative decay~\cite{Besson:1985xw}, following the procedure in Ref.~\cite{Fayet:2007ua}, the present data suggest a very small value for $\cos^{2} \theta$.  However a quantitative estimate cannot be made because only upper limits for the $\Upsilon$ branching fraction are available.  Further measurements of $\Upsilon$ radiative decay to invisible states are warranted.

In summary, the branching fraction for $J/\psi$ radiative decay to invisible particles has been measured as a function of particle mass.  No statistically significant strength was observed for narrow states with mass less than 960 MeV/$c^2$.  The resulting upper limits place new constraints on some theoretical models that include a very light neutralino dark-matter candidate.

% CURRENT acknowledgements go here...
% download from the CLEO website 
% https://wiki.lepp.cornell.edu/lepp/bin/view/CLEO/Private/AC/JournalAcknowledgementsCurrent

\begin{acknowledgements}
We gratefully acknowledge the effort of the CESR staff in providing us with excellent luminosity and running conditions.  D.~Cronin-Hennessy thanks the A.P.~Sloan Foundation. This work was supported by the National Science Foundation, the U.S. Department of Energy, the Natural Sciences and Engineering Research Council of Canada, and the U.K. Science and Technology Facilities Council.
\end{acknowledgements}

\bibliography{clns-2058}        

\end{document}